# Superconductivity in Ternary Zirconium Telluride $Zr_6MTe_2$ with $3d$ Transition Metals

Haruka Matsumoto[1], Youichi Yamakawa[2], Ryutaro Okuma[1],
Daisuke Nishio-Hamane[1], and Yoshihiko Okamoto[1,*]

[1]*Institute for Solid State Physics, University of Tokyo, Kashiwa 277-8581, Japan*
[2]*Department of Physics, Nagoya University, Nagoya 464-8602, Japan*

We report the synthesis, electronic properties, and electronic states of $Zr_6MTe_2$ (M = Cr, Mn, Fe, and Co), which is isostructural to a recently discovered superconductor family $Sc_6MTe_2$. Based on the electrical resistivity and heat capacity data measured at low temperatures, $Zr_6FeTe_2$ is found to show bulk superconductivity below $T_c$ = 0.76 K. $Zr_6CoTe_2$ also exhibited zero resistivity due to superconductivity below 0.13 K. In contrast, $Zr_{6+\delta}Mn_{1-\delta}Te_2$ does not show superconductivity but instead exhibits strong magnetism, which most likely prevents the formation of superconductivity in this material. The electronic properties and electronic states of $Zr_6MTe_2$ are discussed in comparison with those of $Sc_6MTe_2$.

Transition metal tellurides comprise many interesting superconductors as a result of their unique crystal structures, which arise from the chemical bonding of tellurium atoms [1–6]. Recently, ternary scandium tellurides with the formula $Sc_6MTe_2$ were found to be a new $d$-electron superconductor family that exhibits superconductivity in various cases of transition metal elements (M = Fe, Co, Ni, Ru, Rh, Os, and Ir) [7]. The highest critical temperature of $T_c$ = 4.7 K is realized in M = Fe and the $T_c$ decreases in the order of M = Fe, Co, and Ni. $Sc_6MTe_2$ materials with M = $4d$ and $5d$ transition metals show lower $T_c$ of ~2 K. According to first-principles calculations, the electronic states of $Sc_6FeTe_2$ at the Fermi energy $E_F$ mainly consist of both Sc and Fe $3d$ orbitals [7]. In other M cases, the contribution of M $d$ orbitals is less significant than that of Fe $3d$ orbitals in $Sc_6FeTe_2$, suggesting that the $3d$ electrons of Fe atoms play an important role in realizing the highest $T_c$ in $Sc_6FeTe_2$. In contrast, $Sc_6MnTe_2$, where the Mn $3d$ electrons significantly contribute to the electronic states at $E_F$ same as in $Sc_6FeTe_2$, does not show superconductivity, which is probably due to the strong magnetism of Mn $3d$ electrons [7]. Thus, $Sc_6MTe_2$ displays a characteristic M dependence, but it is unknown what kind of electronic properties appears when scandium is replaced by other elements.

Zirconium is a $4d$ transition metal element with one more valence electron than scandium. In this letter, we focus on $Zr_6MTe_2$ with M = Cr, Mn, Fe, and Co. Among them, $Zr_6MTe_2$ with M = Mn, Fe, and Co has been synthesized and reported to crystallize in the hexagonal $Zr_6CoAl_2$ type with the non-centrosymmetric $P\text{-}62m$ space group, as shown in Fig. 1(a), same as in the above-mentioned $Sc_6MTe_2$ [8]. In $Zr_6MTe_2$, M and Te atoms are coordinated by the more electropositive Zr atoms. This situation, in which the transition metal atoms are surrounded by an electropositive element, is the same as in $Sc_6MTe_2$, as well as in the recently discovered superconductors $La_2IRu_2$ and $La_2IOs_2$ [9,10]. This is in contrast to the superconductors ZrRuP and ScIrP with an $Fe_2P$-based crystal structure like $Zr_6MTe_2$, where Ru/Ir atoms are coordinated by phosphorous atoms [11–13]. There have been no reports of the electronic properties of $Zr_6MTe_2$ thus far [8,14,15]. Herein, we find that $Zr_6FeTe_2$ is a bulk superconductor with $T_c$ = 0.76 K. $Zr_6CoTe_2$ also shows zero resistivity due to superconductivity at 0.13 K. Although these $T_c$ values are much lower than those of $Sc_6MTe_2$, $Zr_6MTe_2$ and $Sc_6MTe_2$ share the fact that the highest $T_c$ is realized for M = Fe.

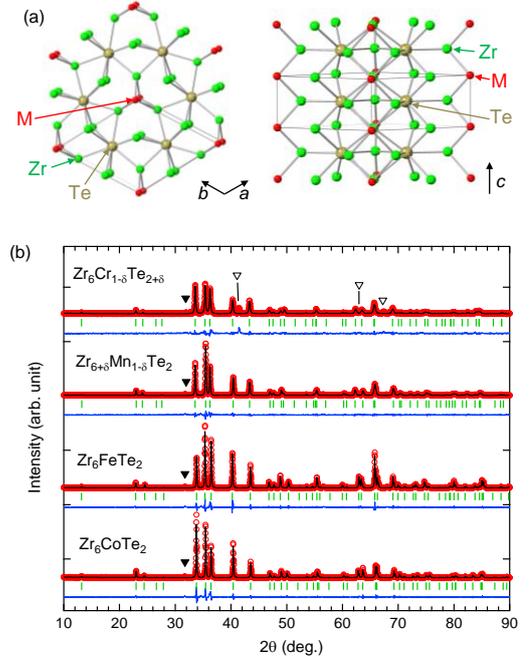

Figure 1. (a) Crystal structure of $Zr_6MTe_2$ (M = Cr, Mn, Fe, and Co) viewed along (left) and perpendicular (right) to the $c$ axis. The solid line indicates the unit cell. (b) Powder XRD patterns of the $Zr_6Cr_{1-\delta}Te_{2+\delta}$, $Zr_{6+\delta}Mn_{1-\delta}Te_2$, $Zr_6FeTe_2$, and $Zr_6CoTe_2$ polycrystalline samples measured at room temperature. Red open circles, black lines, blue lines, and green bars indicate observed diffraction patterns, Le Bail fits, residuals, and Bragg peak positions, respectively. Filled and open triangles indicate the Bragg peaks of impurity phases of Zr and $Cr_2Zr$, respectively. Peak indices are given using hexagonal unit cells with lattice constants of $a$ = 7.7494(3), 7.7278(2), 7.75703(7), and 7.73041(8) Å and $c$ = 3.6766(2), 3.6874(2), 3.62543(7), and 3.64201(8) Å and the agreement factors are $R_p$ = 16.34%, 13.43%, 14.22%, and 16.37% and $R_{wp}$ = 22.56%, 19.23%, 19.75%, and 21.72% for $Zr_6Cr_{1-\delta}Te_{2+\delta}$, $Zr_{6+\delta}Mn_{1-\delta}Te_2$, $Zr_6FeTe_2$, and $Zr_6CoTe_2$, respectively.

$Zr_6MTe_2$ (M = Cr, Mn, Fe, and Co) polycrystalline samples were synthesized by the arc-melting of Zr shot (99.9% and Hf < 50 ppm, RARE METALLIC) and M and Te (99.99%, RARE METALLIC) powders. Cr (99.99%, RARE METALLIC), Mn (99.98%, RARE METALLIC), Fe (99.5%, RARE METALLIC), and Co (99.99%, Wako Pure Chemical Corp.) powders were used for M. First, Zr chips, M powder, and Te powder were weighed in a 6:1:2 molar ratio. For M = Cr and Mn, a 50% excess of Cr/Mn powder was added. The M and Te powders were then mixed and pressed into a pellet. The Zr chips and the pellet were placed on a water-cooled copper hearth and arc melted under an Ar atmosphere. The obtained buttons were subsequently inverted and arc melted several times to promote homogenization. For $Zr_6CoTe_2$, the obtained button was annealed in an evacuated quartz tube at 1123 K for 72 h, which was quenched to room temperature.

Powder X-ray diffraction measurements were performed on a Bragg Brentano diffractometer RINT-2000 diffractometer (RIGAKU) using Cu Kα radiation at room temperature in the 2θ range between 5 and 90° with a step of 0.02°. The data were analyzed by Le Bail method using JANA2006 [16]. As shown in Fig. 1(b), $Zr_6MTe_2$ with a $Zr_6CoAl_2$-type crystal structure was obtained as the main phase. Each sample contained a trace of Zr and for M = Cr, a small amount of an impurity phase $Cr_2Zr$ was detected. The refined lattice parameters were $a$ = 7.7494(3), 7.7278(2), 7.75703(7), and 7.73041(8) Å and $c$ = 3.6766(2), 3.6874(2), 3.62543(7), and 3.64201(8) Å for M = Cr, Mn, Fe, and Co, respectively. The $a$ and $c$ values for Mn, Fe, and Co are consistent with those reported in a previous study [8]. Chemical analyses were conducted using scanning electron microscopy (SEM; JEOL IT-100) equipped with energy dispersive X-ray spectroscopy (EDX; 15 kV, 0.8 nA, 1 μm beam diameter). The ZAF method was used for data correction, and the standards used were pure metals of respective elements. The chemical compositions are estimated to be $Zr_{5.95(4)}Cr_{0.85(2)}Te_{2.20(5)}$, $Zr_{6.22(3)}Mn_{0.73(3)}Te_{2.05(3)}$, $Zr_{5.96(2)}Fe_{0.93(4)}Te_{2.11(3)}$, and $Zr_{5.92(2)}Co_{0.98(3)}Te_{2.10(3)}$, where the total number of atoms in each chemical formula is fixed to be nine, for M = Cr, Mn, Fe, and Co, respectively. This result indicates that the samples for M = Cr and Mn contain a considerable amount of defects at the M site. It is suggested that more than 10% of Cr sites are occupied by Te and more than 20% of Mn sites are occupied by Zr for M = Cr and Mn, respectively. Therefore, in the following, the chemical formulae for M = Cr and Mn are expressed as $Zr_6Cr_{1-\delta}Te_{2+\delta}$ and $Zr_{6+\delta}Mn_{1-\delta}Te_2$, respectively. In contrast, EDX results for M = Fe and Co indicate that the chemical compositions for Fe and Co are more stoichiometric than those for Cr and Mn, although the presence of a slight nonstoichiometry is suggested. Considering a small degree of nonstoichiometry and the 6:1:2 nominal composition ratio of Zr, Fe/Co, and Te for the sample preparation, the chemical formulae for M = Fe and Co are expressed as $Zr_6FeTe_2$ and $Zr_6CoTe_2$.

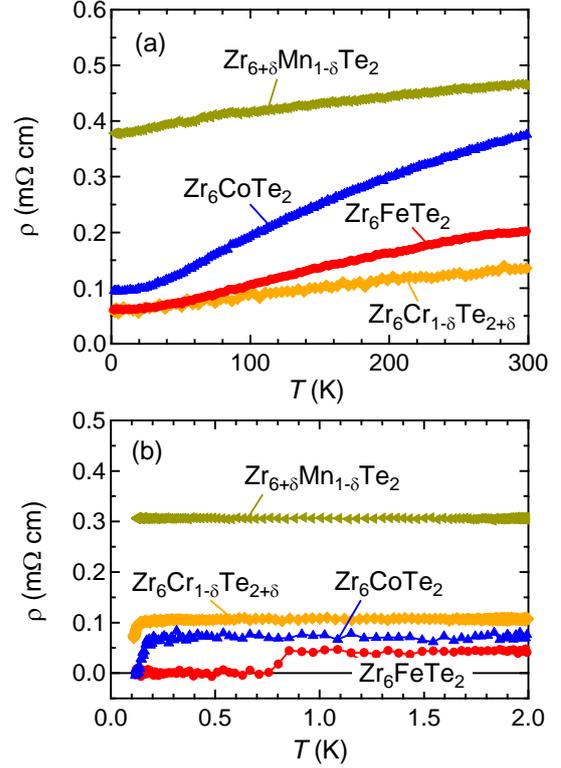

Figure 2. Temperature dependence of electrical resistivity of $Zr_6Cr_{1-\delta}Te_{2+\delta}$, $Zr_{6+\delta}Mn_{1-\delta}Te_2$, $Zr_6FeTe_2$, and $Zr_6CoTe_2$ polycrystalline samples measured above (a) 1.8 K and (b) 0.1 K.

Electrical resistivity and heat capacity measurements were performed using a Physical Property Measurement System (Quantum Design). Electrical resistivity measurements down to 0.1 K were performed using an adiabatic demagnetization refrigerator. Heat capacity measurements down to 0.5 K were performed using a $^3$He refrigerator. Electronic structure calculations for M = Fe were performed using the WIEN2k package in the framework of density functional theory (DFT) based on the full-potential linearized augmented plane wave (FP-LAPW) method [17]. We used the Perdew–Burke–Ernzerhof (PBE) functional, which is one of the most widely used generalized gradient approximation (GGA) based exchange-correlation functionals. Experimentally obtained structural parameters were used for the calculations [8].

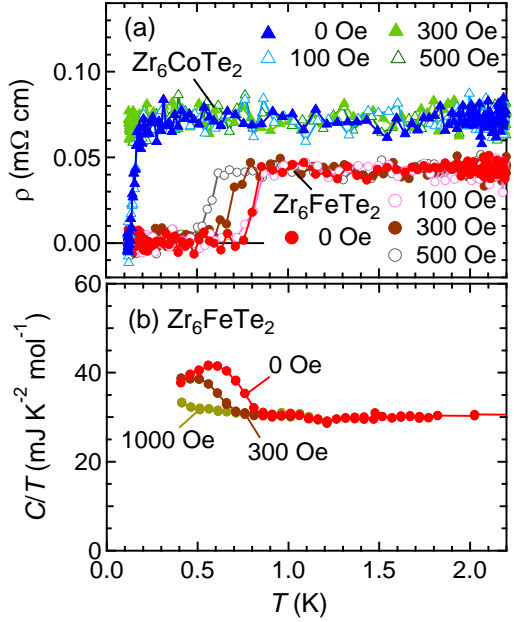

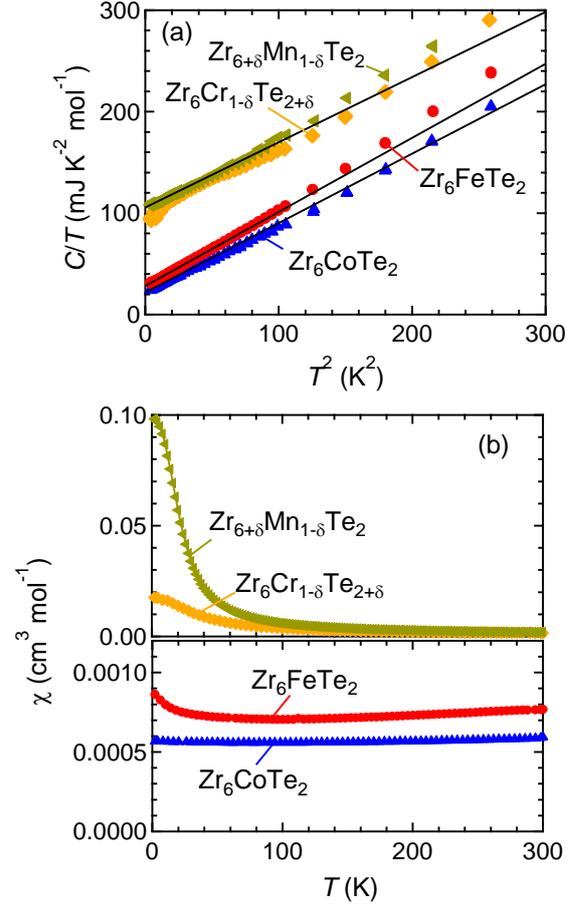

Figure 3. Temperature dependence of (a) electrical resistivity of $Zr_6FeTe_2$ and $Zr_6CoTe_2$ polycrystalline samples measured at magnetic fields of 0, 100, 300, and 500 Oe and (b) heat capacity divided by temperature of a $Zr_6FeTe_2$ polycrystalline sample measured at magnetic fields of 0, 300, and 1000 Oe.

Figure 2 shows the temperature dependence of the electrical resistivity, $\rho$, of the $Zr_6Cr_{1-\delta}Te_{2+\delta}$, $Zr_{6+\delta}Mn_{1-\delta}Te_2$, $Zr_6FeTe_2$, and $Zr_6CoTe_2$ polycrystalline samples. All the samples showed metallic $\rho$ that decreased with decreasing temperature. The residual resistivity ratio, RRR = $\rho_{300K}/\rho_0$, where $\rho_{300K}$ is $\rho$ at 300 K and $\rho_0$ is the residual resistivity, is equal to 2.4, 1.2, 3.4, and 4.0 for $Zr_6Cr_{1-\delta}Te_{2+\delta}$, $Zr_{6+\delta}Mn_{1-\delta}Te_2$, $Zr_6FeTe_2$, and $Zr_6CoTe_2$, respectively. The RRR of $Zr_{6+\delta}Mn_{1-\delta}Te_2$ is considerably smaller than those in the other M cases, which may reflect the strong spin scattering due to the strong magnetism of Mn atoms, as discussed later, and the atomic disorder due to the nonstoichiometry. When the temperature was further decreased, the $\rho$ of $Zr_6FeTe_2$ and $Zr_6CoTe_2$ showed a sharp drop to zero. The onset and zero-resistivity temperatures are 0.85 and 0.76 K for $Zr_6FeTe_2$ and 0.17 and 0.13 K for $Zr_6CoTe_2$, respectively. As shown in Fig. 3(a), these sharp drops of $\rho$ were suppressed by applying magnetic fields, indicative of the superconducting transition. The resistivity of $Zr_6Cr_{1-\delta}Te_{2+\delta}$ also showed a sharp drop at 0.13 K, as shown in Fig. 2(b), but it did not reach zero above the lowest measured temperature of 0.1 K. In contrast, the $\rho$ of $Zr_{6+\delta}Mn_{1-\delta}Te_2$ did not show a strong decrease above 0.1 K.

As shown in Fig. 3(b), the heat capacity divided by temperature, $C/T$, of a $Zr_6FeTe_2$ polycrystalline sample measured at the zero magnetic field strongly increased below 0.8 K with decreasing temperature, followed by a peak at 0.6 K. This temperature corresponds to the sharp drop in the $\rho$ data, indicating that the bulk superconducting transition occurs in $Zr_6FeTe_2$. The peak temperature in the $C/T$ data decreases to 0.5 K at 300 Oe and below 0.4 K at 1000 Oe, consistent with the magnetic field effect on the $\rho$ data shown in Fig. 3(a). Considering the zero-resistivity temperature of 0.76 K and the onset of the heat capacity jump of 0.8 K, the critical temperature of $Zr_6FeTe_2$ was determined to be $T_c$ = 0.76 K. For $Zr_6CoTe_2$, the $T_c$ was determined to be 0.13 K from the zero-resistivity temperature.

Figure 4. (a) Heat capacity divided by temperature of $Zr_6Cr_{1-\delta}Te_{2+\delta}$, $Zr_{6+\delta}Mn_{1-\delta}Te_2$, $Zr_6FeTe_2$, and $Zr_6CoTe_2$ polycrystalline samples as a function of $T^2$. The solid lines show results of the linear fits between 1.9 and 5 K for each sample. (b) Temperature dependence of magnetic susceptibility of $Zr_6Cr_{1-\delta}Te_{2+\delta}$, $Zr_{6+\delta}Mn_{1-\delta}Te_2$, $Zr_6FeTe_2$, and $Zr_6CoTe_2$ polycrystalline samples. The data for $Zr_6Cr_{1-\delta}Te_{2+\delta}$, $Zr_6FeTe_2$, and $Zr_6CoTe_2$ were measured at a magnetic field of $7 \times 10^4$ Oe, whereas those for $Zr_{6+\delta}Mn_{1-\delta}Te_2$ were measured at a magnetic field of $1 \times 10^4$ Oe.

We now discuss the electronic properties of the normal state. Figure 4(a) shows the $C/T$ versus $T^2$ plot for $Zr_6Cr_{1-\delta}Te_{2+\delta}$, $Zr_{6+\delta}Mn_{1-\delta}Te_2$, $Zr_6FeTe_2$, and $Zr_6CoTe_2$ polycrystalline samples in the normal state. The data for $Zr_{6+\delta}Mn_{1-\delta}Te_2$, $Zr_6FeTe_2$, and $Zr_6CoTe_2$ showed a linear behavior at the lowest temperature, whereas those of $Zr_6Cr_{1-\delta}Te_{2+\delta}$ is concave downward below ~6 K. The solid lines on the $Zr_{6+\delta}Mn_{1-\delta}Te_2$, $Zr_6FeTe_2$, and $Zr_6CoTe_2$ data show the fitting result to the equation $C/T = AT^2 + \gamma$, where $A$ and $\gamma$ represent the coefficient of $T^3$ term of the lattice heat capacity and the Sommerfeld coefficient, respectively. The $A$ and $\gamma$ for $Zr_{6+\delta}Mn_{1-\delta}Te_2$, $Zr_6FeTe_2$, and $Zr_6CoTe_2$ were estimated to be $A$ = 0.642(9), 0.731(5), and 0.684(4) mJ K$^{-4}$ mol$^{-1}$ and 105.6(1), 28.10(7), and 21.71(6) mJ K$^{-2}$ mol$^{-1}$, respectively. $Zr_{6+\delta}Mn_{1-\delta}Te_2$ showed a much larger $\gamma$ than those of $Zr_6FeTe_2$ and $Zr_6CoTe_2$. However, considering the strong temperature dependence of magnetic susceptibility $\chi$ shown in Fig. 4(b), it is unclear to what extent the observed large $\gamma$ reflects the electronic density of states (DOS) at $E_F$, because the spin entropy may contribute to the heat capacity at low temperatures. In fact, $Zr_6Cr_{1-\delta}Te_{2+\delta}$ also exhibited strongly temperature-dependent $\chi$, and the spin entropy most likely contributes to the concave downward behavior of $C/T$ at low temperatures. In contrast, $Zr_6FeTe_2$ and $Zr_6CoTe_2$ exhibited an almost temperature independent $\chi$, which can be understood by the Pauli paramagnetism. Considering that there is a more pronounced nonstoichiometry in $Zr_6Cr_{1-\delta}Te_{2+\delta}$ and $Zr_{6+\delta}Mn_{1-\delta}Te_2$ compared to $Zr_6FeTe_2$ and $Zr_6CoTe_2$, the strong increase of $\chi$ at low temperatures in $Zr_6Cr_{1-\delta}Te_{2+\delta}$ and $Zr_{6+\delta}Mn_{1-\delta}Te_2$ is due to the disorder in the magnetic-element sites. The Pauli paramagnetic susceptibility $\chi_{Pauli}$ of $Zr_6FeTe_2$ and $Zr_6CoTe_2$ is $\chi_{Pauli}$ = 1.1 × 10$^{-3}$ and 7.9 × 10$^{-4}$ cm$^3$ mol$^{-1}$, respectively, estimated by subtracting the diamagnetic contribution of core electrons $\chi_{dia}$ = −2.3 × 10$^{-4}$ and −2.2 × 10$^{-4}$ cm$^3$ mol$^{-1}$ from the $\chi$ data at the lowest measured temperature of 2 K [18,19]. These $\chi_{Pauli}$ and $\gamma$ give the Wilson ratio $R_W = (\pi^2 k_B^2/3\mu_B^2)(\chi_{Pauli}/\gamma)$ of 2.8 and 2.6 for M = Fe and Co, respectively. Considering the ambiguity in the estimation of $\chi_{Pauli}$, these $R_W$ values are not far outside those for Pauli paramagnetic metals, suggesting that the observed $\gamma$ values of $Zr_6FeTe_2$ and $Zr_6CoTe_2$ correspond to the DOS at the Fermi energy.

Figure 5(a) shows the electronic band structure and DOS of $Zr_6FeTe_2$ calculated with spin–orbit coupling. The band dispersion of $Zr_6FeTe_2$ is similar to that of $Sc_6FeTe_2$, reflecting the same crystal structure and similar constituent elements [7]. In $Zr_6FeTe_2$, both Zr 4$d$ and Fe 3$d$ orbitals have significant contributions to the electronic states near $E_F$, which is same as in $Sc_6FeTe_2$ (Sc 3$d$ and Fe 3$d$ orbitals for $Sc_6FeTe_2$). However, in $Zr_6FeTe_2$, the electrons occupy considerably higher energy levels than those in $Sc_6FeTe_2$, due to one more valence electron in a Zr atom than in a Sc atom. There is a ~0.5 eV difference between the Fermi levels of $Zr_6FeTe_2$ and $Sc_6FeTe_2$, assuming the rigid-band structure. As a result, these materials have quite different Fermi surfaces. One of the major differences is a spherical hole surface surrounding the $\Gamma$ point, which commonly exists in all $Sc_6MTe_2$ [7] and is absent in $Zr_6FeTe_2$. Nevertheless, the difference in DOS at the Fermi energy $D(E_F)$ between $Zr_6FeTe_2$ and $Sc_6FeTe_2$ is not large; $D(E_F)$ = 8.0 and 9.3 eV$^{-1}$ for $Zr_6FeTe_2$ and $Sc_6FeTe_2$, which yield calculated Sommerfeld coefficient of $\gamma_{band}$ = 19 and 22 mJ K$^{-2}$ mol$^{-1}$, respectively.

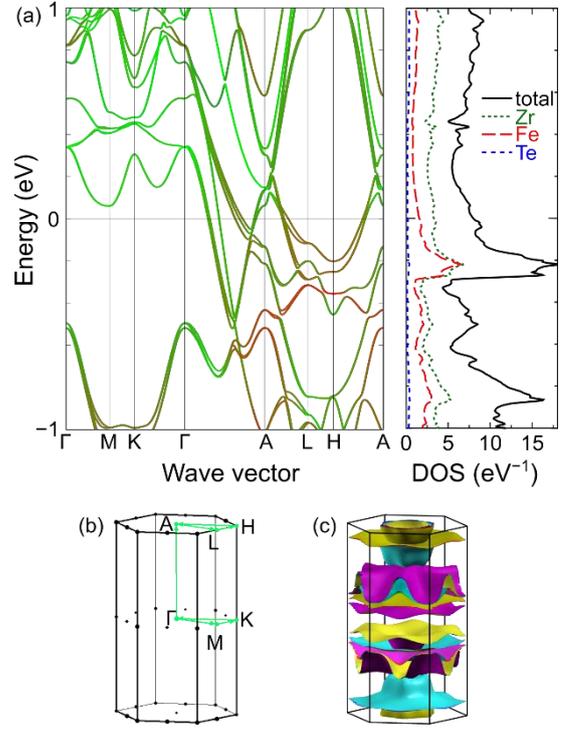

Figure 5. Electronic structure of $Zr_6FeTe_2$. (a) Electronic band structure (left) and total and partial electronic DOS (right) calculated with spin–orbit coupling. The Fermi energy is set to 0 eV. (b) First Brillouin zone. (c) Fermi surfaces calculated without spin–orbit coupling.

In $Zr_6FeTe_2$, the experimentally observed $\gamma$ = 28 mJ K$^{-2}$ mol$^{-1}$ is considerably larger than $\gamma_{band}$, indicating that the $\gamma$ is enhanced by some interaction. However, this enhancement ($\gamma/\gamma_{band}$ = 1.5) is weaker than that in $Sc_6FeTe_2$ ($\gamma/\gamma_{band}$ = 3.3) and rather similar to that in $Sc_6RuTe_2$ with $T_c$ = 1.9 K ($\gamma/\gamma_{band}$ = 1.8) [7]. $Sc_6FeTe_2$ has a large contribution of Fe 3$d$ electrons at $E_F$, whereas that of Ru 4$d$ electrons in $Sc_6RuTe_2$ is small, suggesting that the Fe 3$d$ electrons play an important role in the strong enhancement of $\gamma$ in $Sc_6FeTe_2$. However, the result of $Zr_6FeTe_2$ suggests that the large contribution of Fe 3$d$ electrons at $E_F$ is not sufficient for realizing the strong enhancement of $\gamma$, which may result in the much lower $T_c$ in $Zr_6FeTe_2$ than that in $Sc_6FeTe_2$. The large differences in the $T_c$ values and enhancement of $\gamma$ between $Sc_6FeTe_2$ and $Zr_6FeTe_2$ suggest that $\gamma$ and $T_c$ depend on the different points of these compounds. One is the difference between Sc and Zr. In both compounds, there is a large contribution of Sc 3$d$ or Zr 4$d$ electrons as well as Fe 3$d$ electrons at $E_F$. Sc is a lighter element than Zr, which can give rise to a stronger enhancement of $\gamma$ through electron–phonon interaction. Another point is the different shapes of their Fermi surfaces. As mentioned above, $Sc_6FeTe_2$ has a spherical hole surface surrounding the $\Gamma$ point, which is absent in $Zr_6FeTe_2$. In $Sc_6MTe_2$, this hole surface commonly exists in all M cases, which may be related to the fact that $Sc_6MTe_2$ exhibits superconductivity with similar $T_c$ of 2–5 K for as many as the seven M elements.

Although the $T_c$ values are thus significantly different, $Zr_6MTe_2$ and $Sc_6MTe_2$ showed similar M dependence. The highest $T_c$ is realized in M = Fe, and the $T_c$ value of M = Co is lower. Superconductivity does not appear in M = Mn, probably due to the strong magnetism of Mn atoms. In $Zr_{6+\delta}Mn_{1-\delta}Te_2$, as discussed above, this strong magnetism may be caused by the disorder in the Mn site. At present, it is not entirely clear why $Zr_6MTe_2$ and $Sc_6MTe_2$ exhibit such a common $M$ dependence, even though the shapes of their Fermi surfaces and the $T_c$ values are considerably different. It is hoped that a systematic experimental study on the electronic states will be conducted to clarify the M dependence in $Zr_6MTe_2$.

In summary, $Zr_6FeTe_2$ is found to be a bulk superconductor with $T_c$ = 0.76 K by electrical resistivity and heat capacity measurements on the polycrystalline samples. In addition, $Zr_6CoTe_2$ showed zero resistivity due to superconductivity at $T_c$ = 0.13 K. $Zr_6MTe_2$ showed similar M dependence to $Sc_6MTe_2$ in several points, although the $T_c$ values were significantly different. One is that they exhibit the highest $T_c$ in M = Fe, and the second is that the strong magnetism appears and superconductivity does not appear in M = Mn. According to first-principles calculations, $Zr_6FeTe_2$ and $Sc_6FeTe_2$ have similar $D(E_F)$, although the shapes of their Fermi surfaces are significantly different. It is expected that the entire picture of the large superconductor family including $Zr_6MTe_2$ and $Sc_6MTe_2$ will be elucidated in future studies.

**Acknowledgments**

The authors are grateful to K. Yuchi, K. Moriyama, J. Yamaura, Z. Hiroi, and Y. Shinoda for helpful discussions. This work was supported by JSPS KAKENHI (Grant Nos. 19H05823, 20H02603, and 23H01831).

*email: yokamoto@issp.u-tokyo.ac.jp